\documentclass[aps,prb,twocolumn,groupedaddress]{revtex4-1}
\usepackage{graphicx}

\usepackage{amsmath}
\usepackage{siunitx}

\newcommand{\snr}{\textrm{SNR}}

\newcommand{\eps}{\varepsilon}
\newcommand{\epstl}{\eps_{TL}}
\newcommand{\elf}{\Im(-1/\eps)}
\newcommand{\elftl}{\Im(-1/\epstl)}
\newcommand{\fig}[1]{Fig.~\ref{fig::#1}}
\newcommand{\Fig}[1]{Figure~\ref{fig::#1}}

\begin{document}

%\preprint{}

\title{Design and Application of a Relativistic Kramers-Kronig Analysis Algorithm}


\author{Alberto Eljarrat}
\email[]{aeljarrat@physik.hu-berlin.de}
\author{Christoph T. Koch}

%\thanks{}
\affiliation{Department of Physics, Humboldt University of Berlin, Newtonstra{\ss}e 15, 12489 Berlin, Germany}


\date{\today}

\begin{abstract}
Low-loss electron energy loss spectroscopy (EELS) in the scanning transmission electron microscope (STEM) probes the valence electron density and relevant optoelectronic properties such as band gap energies and other band structure transitions. The measured spectra can be formulated in a dielectric theory framework, comparable to optical spectroscopies and ab-initio simulations. Moreover, Kramers-Kronig analysis (KKA), an inverse algorithm based on the homonym relations, can be employed for the retrieval of the complex dielectric function (CDF). However, spurious contributions traditionally not considered in this framework typically impact low-loss EELS modifying the spectral shapes and precluding the correct measurement and retrieval of the dielectric information. A relativistic KKA algorithm is able to account for the bulk and surface radiative-loss contributions to low-loss EELS, revealing the correct dielectric properties. Using a synthetic low-loss EELS model, we propose some modifications on the naive implementation of this algorithm that broadens its range of application. The robustness of the algorithm is improved by regularization, appliying previous knowledge about the shape and smoothness of the correction term. Additionally, our efficient numerical integration methodology allows processing hyperspectral datasets in a reasonable amount of time. Harnessing these abilities, we show how simultaneous relativistic KKA processing of several spectra can share information to produce an improved result. 
\end{abstract}

%\keywords{}

\maketitle

\section{Introduction\label{sec::intro}}
Low-loss electron energy loss spectroscopy (EELS) combines the ability to measure dielectric properties with ultimate spatial resolution. This ability complements other experimental and theoretical techniques even beyond electron microscopy. For instance, ab-initio simulation codes based on density functional theory (DFT) are able to calculate related quantities with varying degrees of precision \cite{Hebert2007, Keast2013, Eljarrat2015b}. Moreover, optical spectroscopy techniques also measure some of the (optical) transitions observed in EELS \cite{YuCa2011}. Generally speaking, the theoretical framework in which these techniques are formulated has one relevant quantity in common; a complex dielectric function (CDF), that describes the displacement of bound charges in the material when exposed to exterior electric fields . Being able to measure or calculate this quantity is relevant in several fields, for instance to the characterization of semiconductor materials \cite{Eljarrat2012, Eljarrat2014a}. In this sense, a long standing aim exists to use low-loss EELS to perform standard-free measurement of the dielectric properties of material media covering the optical range and up to the ultra-violet (UV) range, without the use of a synchrotron \cite{Daniels1970, Raether1980, Egerton2011}. 

The study of the dielectric response in low-loss EELS is characterized by the choice of a semi-classical or relativistic framework. In the semi-classical formulation, closed formulas describe the energy-loss spectrum in terms of the CDF \cite{Ritchie1957}. Together with the causality properties of the dielectric response, this formulation has been traditionally used in the Kramers-Kronig analysis (KKA) to retrieve the CDF \cite{Egerton2011, Eljarrat2012, Eljarrat2018}. In most cases a relativistic framework has to be considered to completely describe low-loss EELS, as pointed out in early theoretical and experimental work \cite{Kroger1968, Festenberg1968, Daniels1970}. Bulk and surface radiative-loss modes are only explained in this relativistic formulation. It is possible that the generally poor energy resolution of earlier instruments made the study of this modes relatively irrelevant and prioritized the study of surface-losses that impact the spectra at a higher energy-loss range.

The interest in the relativistic formulation increased with the general availability of electron monochromation in STEM-EELS. Relevant (opto)electronic properties, such as the band gap energy in semiconductor materials, can in principle be measured in low-loss EELS with sub-eV energy resolution \cite{Eljarrat2012, Zhan2018}. However, in common experimental conditions, signals indicating the band gap energy onset may be smeared by bulk radiative-loss contributions (i.e. Cerenkov-losses) \cite{Stoeger2006}, making the interpretation of the data problematic. Some experimental methods have been proposed to circumvent or reduce the impact of these spurious contributions and allow the direct observation of the band gap. Among these, using lower voltages and thinner samples can help overall reducing the intensity of bulk radiative-loss \cite{Erni2008a}. Other experimental methods attempt to suppress this contribution by avoiding the forward scattered electrons, either experimentally \cite{Stoeger2008, Stoeger2008, Stoeger2008a}; or by subtracting spectra acquired with different collection apertures \cite{Zhang2008}. 

Since these methods do not guarantee a complete correction of the of the relativistic and surface effects and are not always feasible, interest is brought to off-line analysis methods. The theoretical framework should include the relativistic (bulk and surface) contributions to low-loss EELS, and produce a correction term in order to reveal the naked material dependent spectral features. It is in principle possible to introduce a relativistic calculation into each iteration of the KKA loop in order to calculate such off-line correction \cite{Stoeger2008a}. However, this method faces the problem of the integration of the relativistic double differential cross section (DDCS). For the semi-classical DDCS, this integration is performed analytically, producing a simple model of inelastic scattering. However, the relativistic DDCS has to be integrated numerically, over a mesh of scattering angles. The computations involved are more costly and, as shown below, there a several pitfalls in this process. Some authors have proposed methods to deal with this issues, e.g. by using simple models of the DF in Silicon \cite{Mkhoyan2007}, or more recently by proposing more sophisticated integration methods for the relativistic DDCS \cite{Meng2018}.

In this work, a relativistic KKA algorithm is presented that is designed with speed and reliability in mind and taking hyperspectral analysis into account. This algorithm is based on the traditional KKA loop, using numerical integration of the relativistic DDCS to produce an iteratively updated correction term. Since this computation is costly and not free of errors, several numerical integration methods are implemented in our algorithm and a comparison in terms of their cost and performance is made. To improve the robustness of rKKA against the inaccuracy of the initial guess and noise related issues, we implement simple regularization of the correction term by bounding and smoothing. Finally, a novel method is proposed to integrate the information from the analysis of hyperspectral datasets in which several spectra from the same region are acquired. In these cases, the estimate of the CDF made at different thicknesses can be averaged, further improving the robustness of the algorithm.

The software presented in this work is implemented in Python using the Hyperspy toolbox \cite{hspy}, and is available to fork on github \cite{aeljarrat_repo_r_eels}. The use of fast numerical integration methods and parallel computing makes it generally useful for the simulation and rKKA of EELS spectra and has been tested both in synthetic and experimental data (experimental results are presented in a different paper). 

\section{Materials and methods \label{sec::mam}}
\subsection{Dielectric response model \label{sec::mam::cdf}}
For linear continuous media, low-loss EELS from a thin-film sample is completely described by the dielectric tensor, $\eps=\eps_{ij}(\textbf{q}, E)$; where $\textbf{q}$ is the scattering vector and $E$ is the energy-loss; and a few experiment-dependent parameters \cite{Egerton2011}. From a macroscopic point of view, this is a complex tensor describing polarization in response to outside electric fields. For a microscopic description of the polarization induced by the electron beam, one has to consider a model of the applied perturbation and the bound charge density together with some approximations. Let us consider the special case of small-angle scattering, dictated by the (longitudinal) Coulomb force and Bloch wave-functions in isotropic media. In this case, the electric susceptibility is given by the product of appropriate transition matrix elements and the valence joint density of states (JDOS) \cite{Eljarrat2018}. Additionally, the $\textbf{q}$ dependence and the tensor nature are dropped and the dielectric response is completely described by a complex dielectric function(CDF), $\eps=\eps_R + i\eps_I$. This formalism is equivalent to applying the random-phase approximation (RPA), that is employed in DFT to simulate dielectric properties; or to the Lindhard model \cite{Lindhard1954}, at the core of many CDF models used to fit optical signals. It is an independent particle approximation and in consequence, many-body effects such as spin exchange or Coulomb correlation are not included. However, this simple description is enough for this work, since the interest is focused in obtaining phenomenological insight into the behavior of low-loss EELS in terms of optical transitions. 

With that aim in mind, let us select a CDF model that is useful for the simulation and KKA of low-loss EELS. We describe the imaginary part of this complex function as the sum of individial susceptibilities, $\chi_j$, for each separate contribution \cite{Egerton2011}. For most materials, the main contributions to low-loss spectra come from single electron transitions and a strong plasmon resonance peak. Among the former, the band gap energy onset is perhaps the most relevant feature in semiconductor and dielectric materials. To model a semiconductor featuring a direct band gap transition, the Tauc JDOS model is a natural choice for the susceptibility \cite{YuCa2011}, 

\begin{equation}
 \chi_g = \dfrac{f_g^2 \sqrt{E-E_g}}{E^2} H(E-E_g) 
 \label{eq::01}
\end{equation}

Where $E_g$ is the band gap energy, $f_g$ is proportional to the transition oscillator strength and $H$ is the Heaviside step function. For the plasma oscillation, perhaps the simplest model is the Lorenz oscillator. an hybrid Tauc-Lorenz (TL) model is used in the analysis of ellipsometric data \cite{Jellison1998}, In order to account for the shift of the plasmon resonance induced by the band gap transition,  

\begin{equation}
 \chi_P = \dfrac{f_P^2}{E} \dfrac{ E_P \Gamma_P (E-E_g)^2 } {(E^2-E_P^2)^2+E^2 \Gamma^2} H(E-E_g)
 \label{eq::02}
\end{equation}

Where $E_P$ is the plasmon energy, $\Gamma_P$ is the plasmon width, and $f_P$ is again the resonance strength. The dielectric function model used in the present work is obtained by addition of the two susceptibilities presented above; $\Im(\epstl)=\chi_T+\chi_{TL}$. \Fig{01}a portraits these models for $E_g =$ 1, 3 and 5 eV, in gray-filled areas. Additionally, the absorption is null below the band gap and the absorption decays with an inverse cubic dependence or faster for large energy-loss.

Note that the full complex $\epstl$ (not shown in the figure) is needed for the dielectric model of EELS simulations. The real part can be obtained from the imaginary part using the Kramers-Kronig transform (KKT);

\begin{equation}
 \epstl = 1 + \textrm{KKT}[\Im(\epstl)] + i\Im(\epstl)
 \label{eq::epstl}
\end{equation}

Because of this formulation, the $\epstl$ model agrees perfectly with the Kramers-Kronig relations, which constitute the basic property that enables KKA. We confirm this fact by; transforming back and forth the real and imaginary parts of the $\epstl$ models; and also simulating semi-classical low-loss EELS spectra and processing them with KKA. In both cases, the original and retrieved dielectric functions agree, indicating that the Kramers-Kronig relations hold for $\epstl$. A word of caution: In turn, we observed that for other dielectric models commonly applied in the study of low-loss EELS, e.g. Drude model \cite{Egerton2011}, some of these conditions may not be fulfilled even when a very broad energy range is considered. In those cases the study of KKA is made difficult since the agreement between the original and recovered DF is not guaranteed.

\begin{figure*}
\includegraphics[width=\linewidth]{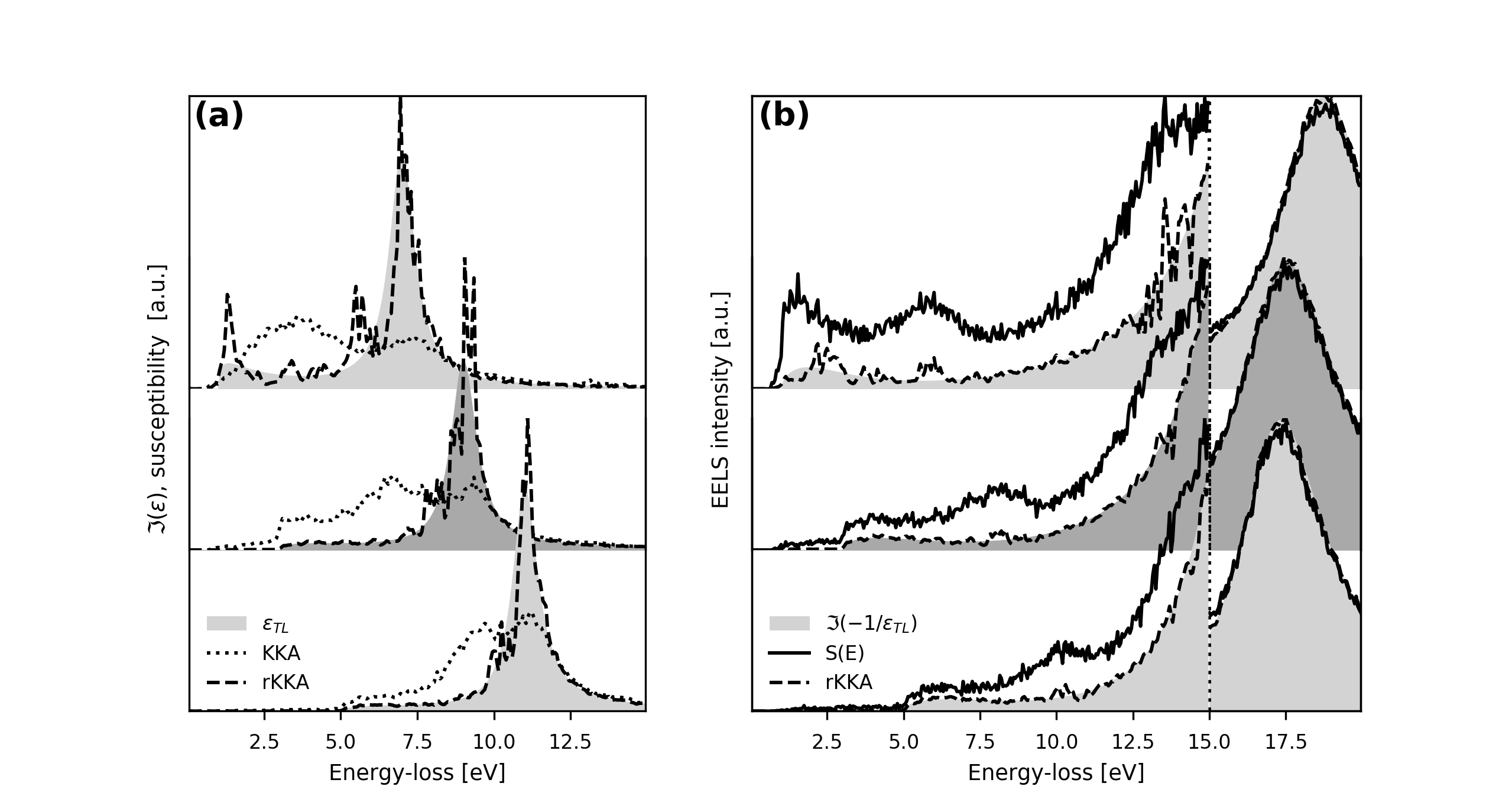}
\caption{ Comparison of the $\epstl$ model simulations and the results from traditional and relativistic KKA, for $t=50$ nm and $E_g=1, 3, 5$ eV, from top to bottom. Panel (a) shows the imaginary part of the models (grey areas) compared to the KKA and rKKA reconstructions (fine pointed and dashed lines, respectively). Panel (b) shows the $\elf$ contribution (grey areas) and the relativistic spectra (solid lines) calculated for these models. A dashed line shows the estimated $\elf$ contribution, after rKKA correction $S_c$ is applied. Details of the simulation and analysis parameters are found in the text. \label{fig::01}}
\end{figure*}

\subsection{Dielectric formulation of EELS \label{sec::mam::eps}}
Single scattering distribution (SSD) energy-loss spectra, $S(E)$, can be calculated for the $\epstl$ models (see for instance \fig{01}). Disregarding plural scattering, these $S(E)$ emulate experimentally obtained low-loss EELS for thin-film samples. These calculations are performed using a formulation of EELS that links dielectric theory and the observed low-loss spectra. This dielectric formulation of EELS considers Maxwell equations, solved for the charge distribution in the bulk and boundaries of the material media. Since the potential can be generally separated into bulk and boundary reflection terms, separate expressions for the bulk and surface DDCS \cite{GarciaDeAbajo2010}. The total spectrum can be found by integrating these DDCS and adding each contribution, $S = S_b+S_s$;

\begin{equation}
 S_{b,s} (E) = \int_0^{\theta_{max}} D_{b, s} (\theta, E) \sin(\theta) d\theta
 \label{eq::03}
\end{equation}

Where $D_{b, s}$ and $S_{b,s}$ are the bulk and surface DDCS and energy-loss spectra, respectively; and $\theta$ is the scattering angle. For both bulk and surface modes, depending if retardation effects are considered in the DDCS models, semi-classical and relativistic contributions to the total spectrum can be identified. 

For the semi-classical case retardation effects are disregarded and analytical integration of the DDCS is possible \cite{Ritchie1957}. Then, closed formulas that model $S(E)$ as a function of $\eps(E)$ can be obtained, taking into account only a few parameters; the incoming beam energy, $E_0$; the transversed material thickness, $t$; and the scattering angle cut-off, $\theta_{max}$.  In this formulation the bulk term is proportional to the inverse of the dielectric function multiplied by a known angular integration term, 

\begin{equation}
 S_b^{ELF}(E) = \frac{2I_0t \elf}{\pi^2a_0m_0v^2} \log \left[1+(\theta_{max}/\theta_E)^2\right] 
 \label{eq::elf}
\end{equation}

Where $I_0$ is the zero-loss intensity, $a_0$ is the Bohr radius, $m_0$ is the electron rest mass, $v$ is the electron speed and $\theta_E = E/(\gamma m_0 v^2)$ is the characteristic scattering angle. Moreover, $\elf$ is also called the energy-loss function (ELF). This ELF produces a contribution that has a fixed shape given by eq.~\ref{eq::elf} and scales linearly with thickness. For samples with thickness above a few tens of nm this contribution always dominates $S(E)$. Examples can be found in \fig{01}, where the spectra in panel (b) correspond to $\elftl$. 

However, a mostly thickness independent surface-loss term from the sample boundaries always exists. A closed expression for this term also exists but will not be reproduced here, the reader is refered elsewhere \cite{Eljarrat2018}. The surface contribution modifies the spectral shape, principally by adding an additional surface-plasmon peak that can be observed and sometimes dominates in very thin specimens. Additionally, $S_s(E)$ has a negative intensity spectral region representing a reduction of the EELS signal. 

A full relativistic description (i.e. including retardation effects) has to be considered when the speed of the fast electrons surpasses that of light in the medium $\eps_R>c^2/v^2$. This is a common case for the analysis of materials in the STEM, because of the high voltages used in the electron beam. In those cases, including retardation effects into the DDCS produces a more intricate model with additional contributions \cite{Kroger1968},

\begin{align}
\begin{split}
      &D(\theta, E) = D_b + D_s = ~ \\
  \\ &\frac{I_0}{\pi^2a_0m_0v^2} \Im\left[\frac{t\mu^2}{\eps^*\varphi^2}-\frac{2\theta^2(\eps^*-\eta^*)^2}{k_0\varphi_0^4\varphi^4}(A+B+C)\right] 
\end{split}
\label{eq::kro}
\end{align}
 
Where $\eps^* = \eps_1 - i\eps_2$ is the complex conjugate of the dielectric function for the specimen, and $\eta^*$, idem for the surroundings (in this work, $\eta^* = 1$ for vacuum). Moreover, eq.~\ref{eq::kro} is given as a function of adimensional terms $\mu$ and $\varphi$ for momentum exchange and the A, B, and C terms representing different surface-loss terms, by surface-plasmon and guided-light modes. To avoid cluttering, the relatively intricate dependence of these terms on $\theta,~E$ and $t$ is not described here, their definitions can be found elsewhere \citep{Egerton2011}.

This complexity poses a difficult challenge to analytically solve eq.~\ref{eq::03}, and to our knowledge there are no available closed formulas for $S(E)$ in the relativistic formulation. Nevertheless, the relativistic DDCS can be numerically integrated (more below). For thin-film samples and comparing to the semi-classical formulation, once the retardation effects are taken into account this means a radical modification of the bulk and surface terms, although the thickness behavior is similar. Bulk radiative-loss excitation is now possible, emitting Cerenkov radiation with an intensity directly proportional to the thickness. Additionally, a variety of boundary coupling effects are observed depending on the interfaces of the material media. 

\begin{figure*}
\includegraphics[width=\linewidth]{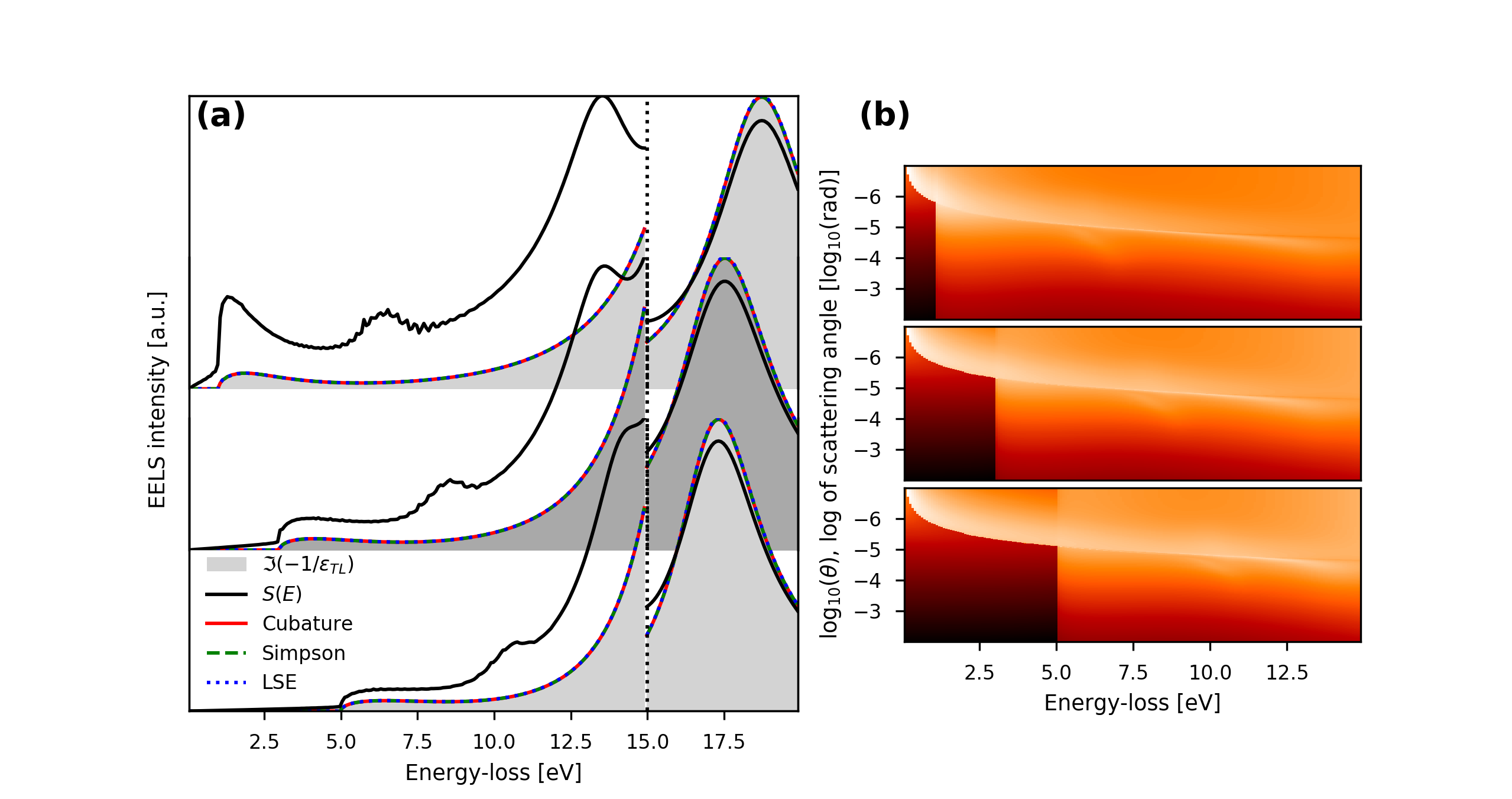}
\caption{ For the same three $\epstl$ models presented in \fig{01}, panel (a) is showing simulated relativistic spectra (solid lines) while their corresponding DDCS log-mesh can be found in panel (b). Panel (a) also compares the analytical $\elf$ contribution to the numerical estimation obtained using cubature, Simpson-rule and LSE trick methods. \label{fig::02}}
\end{figure*}

Figures~\ref{fig::01}b and~\ref{fig::02}a are helpful to understand the importance of bulk and surface, semi-classical or relativistic contributions better. In these panels, the featured $S(E)$ are calculated using the numerically integrated full relativistic model of eq.~\ref{eq::kro} (solid black lines) and compared to the contribution of the bulk semi-relativistic term described by eq.~\ref{eq::elf} (in grey areas). The latter is clearly dominant, however $S(E)$ departs from the shape dictated by $\elftl$ by an the additional peak at around \SI{15}{\electronvolt}, mainly due to the surface plasmon. Additionally, the spectral shape in the lower energy-loss range, close to the band gap energy onset, is radically modified; this time by Cerenkov-loss. Finally, note that a Poisson-distributed random contribution has been added to these spectra to simulate the effect of noise at the detector. 

The relativistic DDCS corresponding to these spectra are also depicted in \fig{02}b.  As advanced, numerical integration is employed to simulate relativistic spectra, making the calculations much more demanding than for the semi-classical model. For this task, a DDCS mesh with one entry for each pair of scattering-angle and energy-loss values is used. From these, a numerical integration routine of choice estimates the angle-integrated SSD. For the relativistic DDCS, the use of a logarithmic mesh (log-mesh) of angles is customary, since accounting for small-angle variations with a linear mesh would require a huge number of entries. The reasons for this are visible in \fig{02}b; principally that radiative-loss modes that appear at very small scattering angles ($\theta\sim$ \SI{}{\micro\radian}); all the while, the spectra are usually acquired with relatively large cut-offs ($\theta_{max}\sim$ \SI{}{\milli\radian}), to increase counting statistics. 

Numerical integration of the DDCS constitutes a slow and error-prone process, which in this work we aim to optimize. The main reason for this is that simple numerical integration algorithms are not useful to integrate the DDCS since we use an irregularly-spaced angular log-mesh, and more sophisticated methods have to be applied. We have performed benchmark tests of integration methods, as illustrated in \fig{02}a, the outcomes of which will be discussed in Sec.\ref{sec::results::ddcs}.

\subsection{Kramers-Kronig analysis \label{sec::mam::kka}}
The KKA algorithm is an inverse algorithm reconciling the dielectric response and EELS models; from EELS measurements it aims to reveal the dielectric properties. In its original formulation, KKA uses the expressions of the semi-classical dielectric formulation to relate the measurements to the ELF. Additionally, surface-loss contributions are measured and suppressed. This algorithm is commonly implemented as an iterative loop with 4 main steps, as depicted in the diagram in \fig{03}. These steps are explained below, without going into unnecessary detail. The basis of this method is also explained at length and including some application examples elsewhere \cite{Egerton2011, Eljarrat2018}. 

As explained above, in the semi-classical approximation $\elf$ can be obtained from normalization of $S_b$, the main contribution to $S(E)$. In order to obtain the normalization factor, knowledge of the sample thickness is necessary. In the cases where this parameter is not accurately known, the Kramers-Kronig sum-rule can be used (\textit{e.g.} in experimental application), obtaining in turn an estimation of the thickness. After normalization the full dielectric function is obtained by application of the Kramers-Kronig transform, $\Re(1/\eps)=1-\textrm{KKT}[\elf]$, and simple algebra. Using fast Fourier transform (FFT), a fast and reliable time-domain method can be applied given that the EELS intensity at high energy-loss decays smoothly \cite{Johnson1975}. Using this procedure at each iteration, $i$, and for each input spectrum $S_i$, a dielectric function, $\eps_i$ is estimated.

Having reached this point (point 3 in \fig{03}), it is important to note that even if the normalization factor is perfectly known the resulting estimate of the ELF contains spurious contributions. In the semi-classical model, these stem from the ignored surface-loss term. Moreover, the dielectric function retrieved after applying the Kramers-Kronig transform is in principle also affected by these contributions. Consequently, the last two steps of the KKA loop are aimed at measuring the spurious contributions present in the original input signal, in order to suppress them from the estimate of $S_b$. Since KKA is formulated in a non-relativistic framework, this contribution is only $S_s$.

In order to estimate $S_s$ the current guess of the CDF, $\eps_i$, together with the same parameters employed for the normalization step are used to calculate a correction term, equivalent to the surface contribution, $S_{s, i}$, of an underlying model of the signal, $I_i(E) = S_{b, i}+S_{s, i}$. 

At the end of each iteration, the correction term is applied as a correction to the original input spectrum, updating the input $S_{i+1}(E)$ used for the next iteration. For obvious reasons, this correction term is termed surface-plasmon estimation. Traditional KKA is fast and reliable, and usually converges after a few iterations \cite{Egerton2011}. A calculation can be considered converged either when the underlying model and the original spectrum are equal, $I_i(E)\simeq S(E)$; or, alternatively, when the correction does not change any more between iterations, $S_{s, i}\simeq S_{s, i-1}$. 

However, traditional KKA neglects relativistic terms and does not perform well when these are included in the input $S(E)$. These contributions are present in real EELS spectra, and the CDF retrieved from the application of KKA to these are known to contain errors \cite{Stoeger2006, Stoeger2008}. Some example results from the application of KKA to relativistic spectra could already be examined in \fig{01}a, with pointed lines. The correspondence between these results and the original $\epstl$ is quite poor, especially below 10 eV and further down in the optical regime.

\begin{figure}
\includegraphics[width=\linewidth]{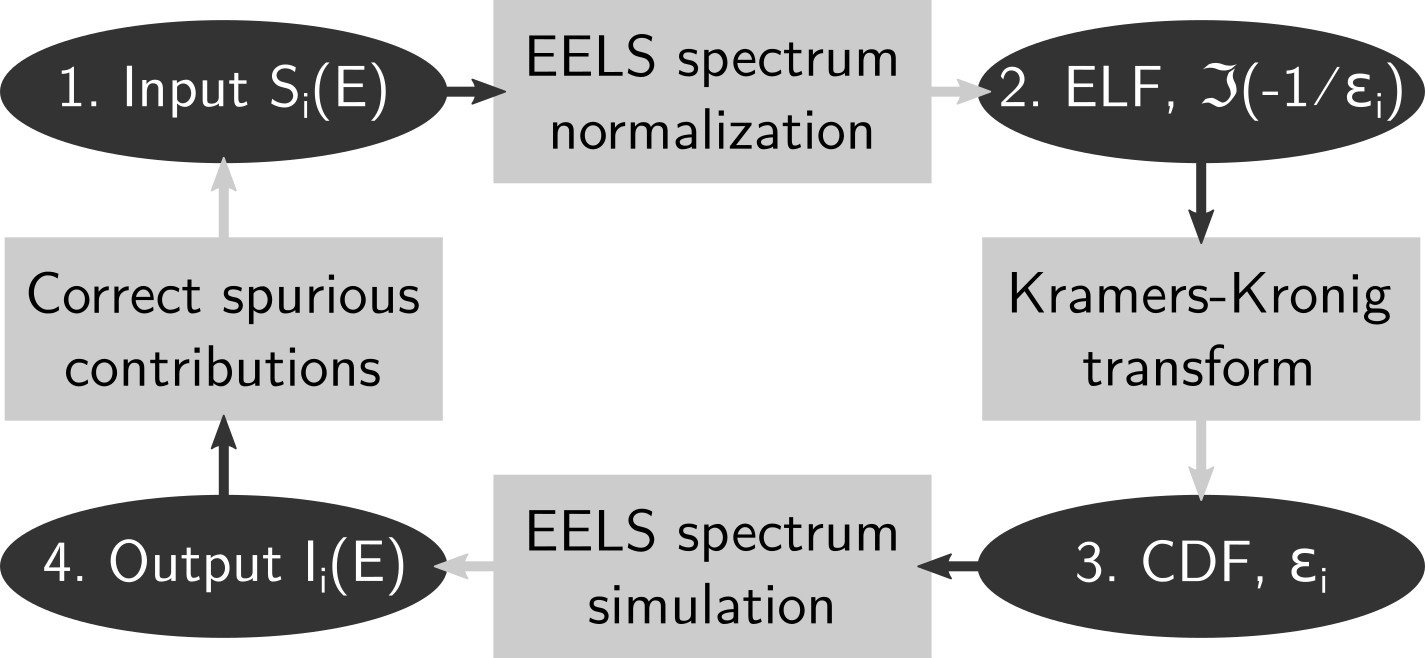}
\caption{Diagram showing the steps of KKA; starting from the normalization of the spectrum to obtain an estimate of the ELF; the Kramers-Kronig transformation to retrieve the CDF; and simulation of the underlying model to correct the spurious contributions. \label{fig::03}}
\end{figure}

\subsubsection{Relativistic Kramers-Kronig analysis \label{sec::mam::kka::kroeger}}
In order to broaden the range of application of traditional KKA, it has been proposed to use the relativistic formulation as the underlying model for the calculation of the correction term \cite{Stoeger2008a}. This constitutes the framework for a relativistic KKA (rKKA), in which a new correction term, $S_c (E)$, contains all contributions except from the bulk non-relativistic term, $S_b^{ELF}(E) = S - S_c$ (see eqs.\ref{eq::elf} and \ref{eq::kro}).

Following these principles and with efficiency in mind, we implement rKKA using a modified DDCS, $D_c(\theta, E)$, calculated as;

\begin{equation}
 D_c(\theta, E) = D(\theta, E) - D_b^{ELF}(\theta, E)
 \label{eq::04}
\end{equation}

Where, $D_b^{ELF}(\theta, E)$ is the DDCS corresponding to the bulk semi-classical term $S_b^{ELF}$. Using this method, the correction can be calculated using a single numerical integration of the DDCS, which is desirable since this computation is costly. 

These are the main ingredients for our rKKA implementation, and even in this basic form, the results are quite good, but not excellent. \Fig{01}a shows the CDF retrieved after applying rKKA (dashed lines). The first thing we notice is that the correspondence between these results and the original $\epstl$ is much better than for traditional KKA, reproducing the dielectric response down to the optical regime. Additionally, since the estimated $S_c$ contains all relativistic contributions, it can be applied to the input spectrum to reveal $S_b^{ELF}$. This procedure is depicted in \fig{01}b, with the resulting spectra (dashed lines) showing good agreement with the theoretical $\elftl$.

However, the retrieved dielectric functions contain some rippling features, obviously artifacts not observed in the original $\epstl$ models and the input $S(E)$ spectra. Examining also these spectra, it is clear that related errors are more important near the band gap onset and at around 15 eV. Cerenkov and surface losses respectively impact these two regions, and the results indicate that the appearance of these rippling features can be related to the incomplete suppression of these spurious contributions. 

The origin of these issues, the effects of which can be observed in other similar works \cite{Meng2018}, is two-fold. Either the numerical integral is completely reliable; or the initial guess of the CDF is too far from the ground truth. Gross errors are indeed apparent in the correction terms in the form of intense peaks that in turn produce spectral regions of negative intensity and high-frequency noise that can ultimately preclude the convergence of the algorithm.

In order to investigate and propose solutions to these issues, we analyze low-loss EELS synthetic data using our own EELS simulation and rKKA algorithms. Computing time is also considered as hyperspectral acquisition methods are widespread and we can use this extended source of information to our advantage (see Sec.\ref{sec::results::average}). Our aim is to be able to treat batches with many spectra at the same time, consequently efficient DDCS integration using several methods and parallel computation are investigated. 

For this purpose, we use $\epstl$ models, with band gap energies between 1 and 6 eV, to calculate semi-classical and relativistic DDCS log-meshes. The parameters of the simulation are fixed to $E_0 =$ \SI{300}{\kilo\electronvolt}, $t =10-$\SI{250}{\nano\meter} and $\theta_{max}=$\SI{10}{\milli\radian}. Numerical integration of the log-meshes is optimized for speed and reliability, from $\theta_{min}=0.1-$\SI{1}{\micro\radian}, and an angular mesh size $N_{\theta}=256-512$. Poisson noise is added to the spectra used as input to the rKKA algorithm, to investigate also the effects of counting statistics. In this sense, simple trial tests indicate that to make the signal-to-noise (SNR) ratio drop appreciable for the spectra with low number of counts, a zero-loss intenstity parameter $I_{0}=1\cdot 10^6~e^-$ is sufficient. 

The rKKA algorithm is initialized using these spectra containing all relativistic contributions. The normalization is performed using the thickness as a known parameter (refractive index normalization is also possible, but not used here). The rKKA loop runs until either convergence or a maximum of 20 iterations are reached. Convergence is indicated by the variation between iterations of the estimated relativistic correction, measured using a weighted test, $\chi^2(S_{s, i}, S_{s, i-1})<5\cdot10^{-4}$. The retrieved CDF and estimated ELF contributions can then be compared with the known ground truth counterparts.

\section{Results \label{sec::results}}
\subsection{Optimization of the relativistic DDCS integration \label{sec::results::ddcs}}
We take into account several numerical integration methods (see \fig{02}a), running benchmark tests against the semi-classic dielectric model to test their reliability. Additionally, computational cost tests where performed measuring the average time spent in the calculation of a relativistic spectrum for datasets with sizes between 8 and 64 spectra (solid lines). In all cases, parallel processing was used in a workstation with 8 CPUs and 32 Gb of RAM.

The considered methods include Gaussian quadrature/cubature, Simpson rule and log-sum-exp (LSE) trick integration. Gaussian quadrature integration is perhaps the most popular solution, already implemented in a freely available Matlab low-loss EELS simulation package \cite{Egerton2011}. Inspired on this solution, we have implemented a faster, multidimensional version using the freely available cubature Python wrapper \cite{castro2015, castro2015a, castro2017}. Simpson-rule method is based on the well-known numerical integration formula, generalized for irregularly-spaced data meshes. This method is implemented using the routine already available in the Hyperspy toolbox \cite{hspy}. Finally, LSE trick is the more straightforward solution of summing though the values in a log-mesh and performing the appropriate change of variables. This method is easily implemented based on the LSE routine available in the scipy package \cite{scipy}.

Our tests indicate that the Simpson-rule method gives a good balance between speed and reliability. It produces an optimum estimate and is the less time-consuming for medium size datasets, scoring between 0.25-0.14 s/spectrum for dataset sizes 4-64 spectra. In both this method and the LSE trick, and for larger datasets, the calculations benefit from cached operations meaning that the speed per spectrum increases. The LSE method is the less time-consuming method for larger datasets (below 0.13 s/spectrum), however, even if over/under-flow errors are taken into account it proves to be the less reliable. The errors are however small, and thus difficult to appreciate in \fig{02}a.

Finally, the cubature/quadrature methods are reliable but also more demanding, additionally requiring interpolation of the data prior to numerical integration. They are the most time-consuming, scoring 2 s/spectrum over all test sizes for the more efficient cubature method. Thus the cubature integration is not practical for performing fast batch calculations with many spectra. It is however useful to run tests when the DDCS angular mesh is being optimized.

The final version of our algorithm incorporates all the three featured methods, apart from the slower quadrature method (included for legacy reasons). Furthermore, routines for the prediction of the angular spread of radiative and non-radiative bulk inelastic scattering have also been incorporated. These are useful for the optimization of the DDCS log-meshes, and have been used together with the efficient Simpson-rule integration method for all the remaining calculations presented ($\theta_{min} =$ \SI{1}{\micro\radian}, Ntheta = 256). Using this optimized integration scheme, the simulation and rKKA processing of hyperspectral datasets with a few hundred spectra in a matter of minutes is possible.

\begin{figure}
\includegraphics[width=\linewidth]{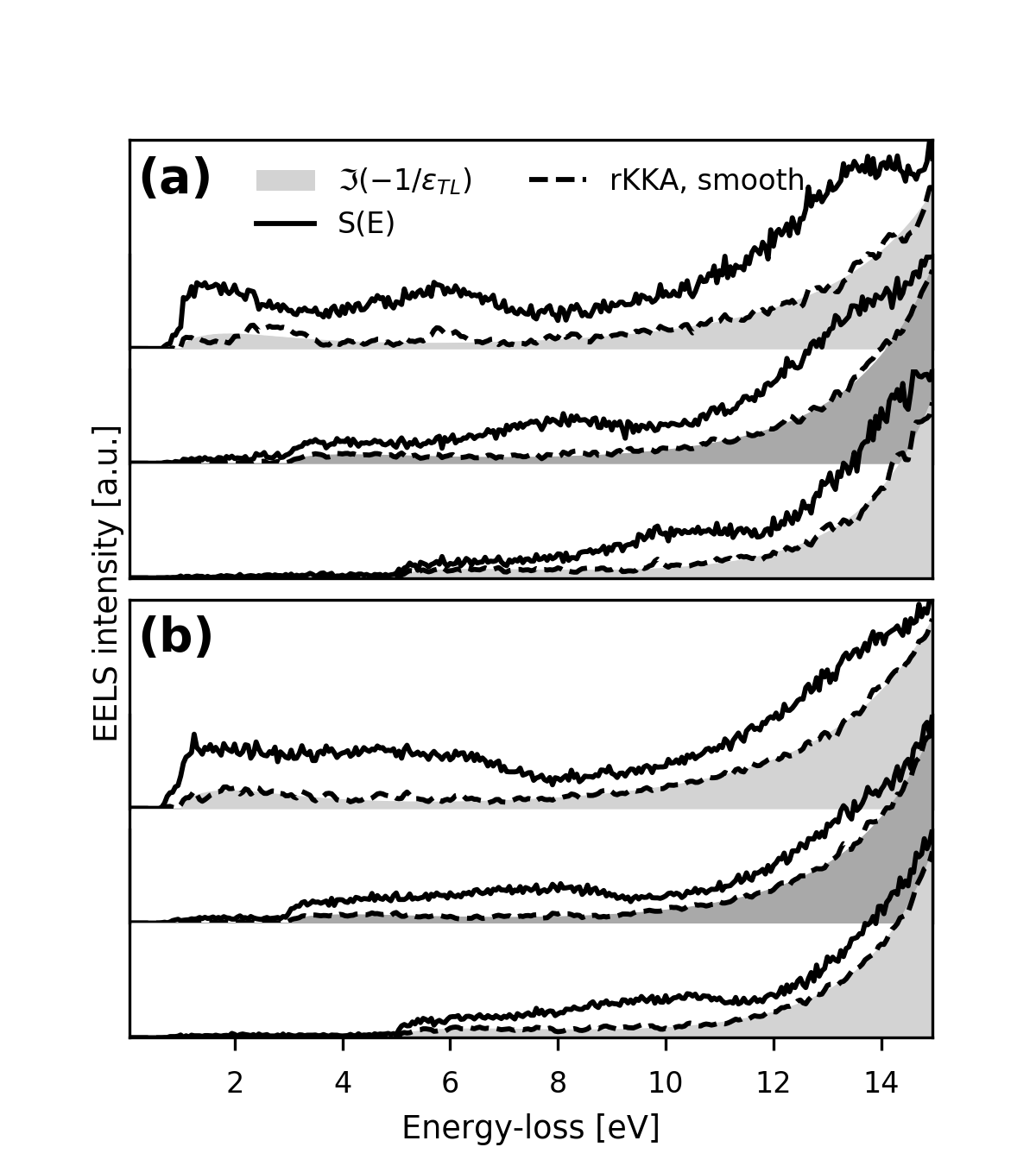}
\caption{For the same three $\epstl$ models presented in \fig{01}, panel (a) and (b) present regularized rKKA results for $t=50$ and 100 nm, respectively.\label{fig::04}}
\end{figure}

\subsection{Relativistic KKA of single spectra \label{sec::results::single}}
In our preliminary results using a naive rKKA implementation (see \fig{01}), we observe rippling errors associated with the incomplete suppression of spurious contributions. Our investigation shows that these issues have a greater impact when surface-loss and Cerenkov-loss terms are relatively intense. Considering also the spectra in \fig{02}a, we observe that for the same thickness the impact of these contributions is greater when the $\epstl$ model with lower band gap energy is employed. The reason is that the model puts a greater oscillator strength into the absorption spectra, and consequently the spectral features are more pronounced. The physical equivalent would be a material that has a larger refractive index, for which the impact of radiative-loss is naturally more important \cite{GarciaDeAbajo2010}.

Our final implementation of rKKA uses regularization of the correction term $S_c$ by bounding and smoothing to improve the reliability of the iterative solutions. In this sense; bounding means that at each iteration the correction values at a given energy $S_c(E')$ are limited to a fixed percentage of the total intensity,

\begin{equation}
 S_c^{bound}(E) = 
 \begin{cases}
    S_c(E'),               & S_c(E') < S(E')*b_{KKA} \\
    S_c(E')b_{KKA}, & \text{otherwise}
 \end{cases}
\end{equation}

Where $b_{KKA}\in(0,1)$ controls the bounding limits; \textit{e.g.} $b_{KKA}=1$ means that the correction can be exactly equal to the input signal but not greater. Following this procedure, smoothing is performed using a Gaussian filter set to a desired energy broadening,

\begin{equation}
 S_c^{gauss} (E) = S_c^{bound} * G(\gamma_{KKA})
\end{equation}

Where the right-hand side denotes convolution with a Gaussian kernel, $G$, with an energy broadening parameter $\gamma_{KKA}$; \textit{e.g.} of a few tens of \SI{}{\electronvolt}.

This approach is equivalent to imposing previous knowledge about the intensity and shape of the correction term. In this sense, a $b_{KKA}<1$ ensures that the intensity of $S_c$ never surpasses the original $S$, introducing regions of negative spectral intensity in the input of the next iteration. Additionally, a moderate smoothing avoids the introduction of high-frequency oscillations which are not suppressed by the iterative reconstruction while preserving the relevant features of $S_c$.

The regularization procedures ensure that the calculated corrections converge and eliminates any rippling features in most cases. More precisely, for all the simulated spectra the solutions above $40-$\SI{50}{\nano\meter} thickness in our simulations have an excellent agreement with the original TL-DF and expected correction terms. 

\Fig{04} showcases the correction results obtained using rKKA for thicknesses of 50 and 100 nm. These corrections are bound to the input spectra by $b_{KKA}=0.99$, and high-frequency oscillatory components above $\gamma_{KKA}=$\SI{0.2}{\electronvolt} are dampened. Examination of the spectra obtained after the application of these correction terms (dashed lines) confirms that the rippling features are largely removed. 

Only the result obtained for \SI{1}{\electronvolt} band gap energy and \SI{50}{\nano\meter} thickness diverge noticeably from the expected bulk semi-classic contribution in the band gap energy onset region. In contrast, these issues are not affecting the result for $t=$\SI{100}{\nano\meter}, although the intensity of Cerenkov-loss increases with thickness. The reason for this discrepancy is certainly the worst counting statistics for a thinner sample, a problem which is exacerbated in our synthetic datasets by the addition of Poisson noise.

In the presented cases, the choice of bounding and smoothing parameters is not a great issue, since we have noticed that the values described above work well in most cases. In experimental application, smoothing can be adjusted attending to the energy resolution. Special cases in which initially running some iterations with a more limiting bounding, or greater smoothing, can also be explored. Note however that depending on the chosen parameters the number of iterations employed to reach convergence can change. In the worst cases, the predicted and original SSD may not converge in cases were the correction ground truth is above the selected percentage or the smoothing is too large and relevant features are distorted or removed.

For relatively thinner samples, however, gross errors are introduced that are only attenuated but can not be completely corrected using only the presented regularization methodology. An increasing impact of these artifacts as the simulated thickness decreases can be observed in \fig{05}, red lines. We determine that the origin of these issues is not the noise-response of the correction calculation, but the inadequacy of the initial guess for the DF; see steps 2 and 3 in \fig{03}, respectively. In this sense, thinner regions have a relatively larger impact of surface losses. To eliminate the errors, it is desirable to improve the initial guess of the CDF, prior to the estimation of $S_c$. In the following section, we devise a methodology to do so, incorporating the information from hyperspectral datasets.

\begin{figure}
\includegraphics[width=\linewidth]{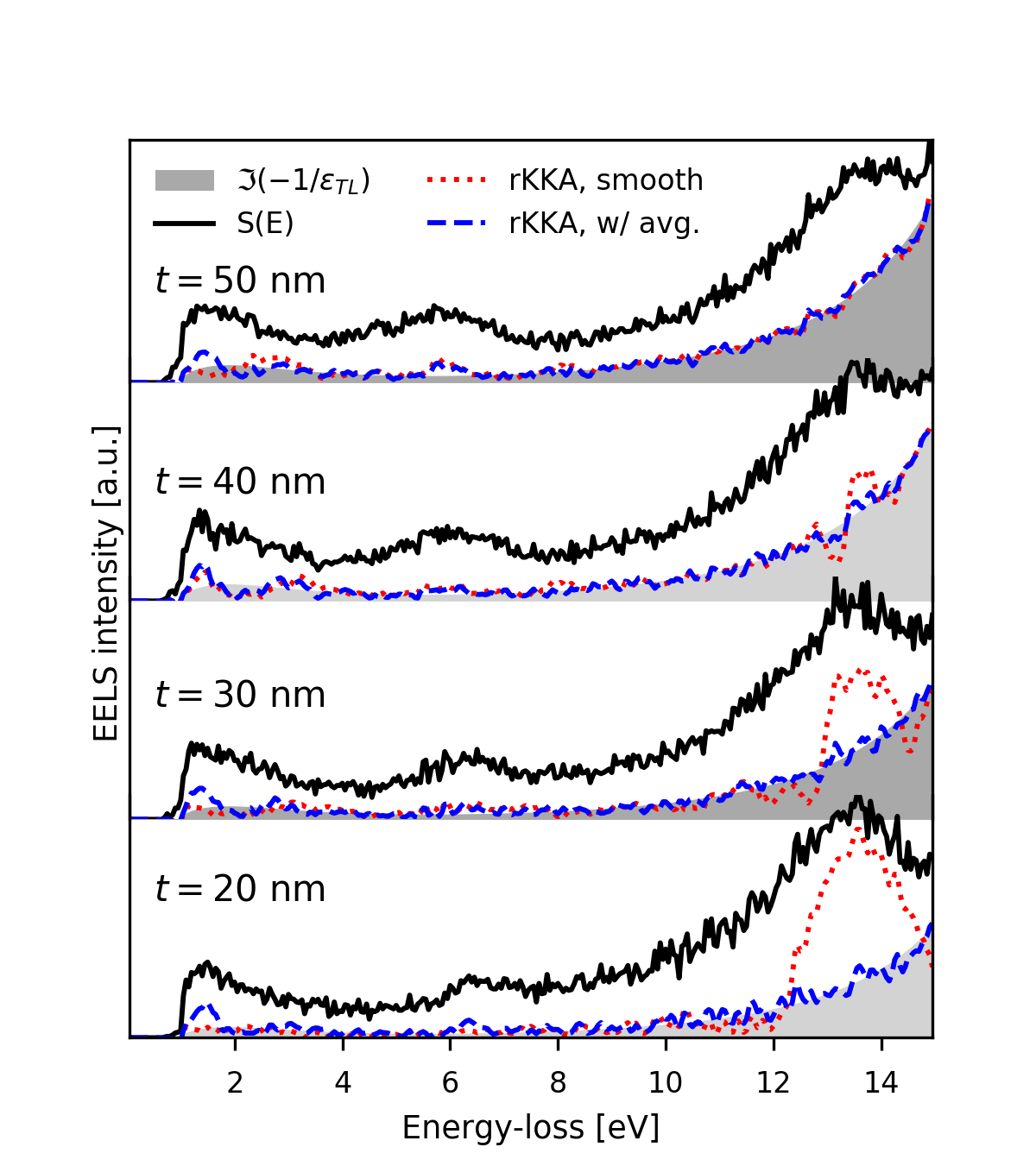}
\caption{For the same $\epstl$ models with $E_g=1$ eV presented in \fig{01}, this figure presents regularized and average rKKA results for several thickness values, with red and blue dashed lines, respectively. Details of the simulation and analysis parameters are found in the text. \label{fig::05}}
\end{figure}

\subsection{Relativistic KKA of an EELS-SL \label{sec::results::average}}
Hyperspectral acquisition modes, in which many spectra from the same material region are acquired and analyzed, is a normal practice in experimental applications. For isotropic media, we can make the assumption that the only difference between each spectra acquired from the same material region is the material thickness transversed by the electron beam. In this case, the dielectric properties originating the EELS are the same for each spectra and contained in a single $\eps(E)$ function. This situation is easily replicated in our simulations, that generate synthetic datasets equivalent to hyperspectral line profiles, commonly known as EELS spectrum-lines (SL). In these EELS-SL a linear thickness gradient exists between the spectra, and they are affected differently by spurious contributions; see \fig{05}, solid black lines.

Under the above assumption, it is therefore natural to use the same guess for the CDF to model each single spectrum in our synthetic EELS-SL and calculate $S_c$. When this is done, the $S$ and $S_c$ in the EELS-SL are still different to each other, since they are calculated for their corresponding thickness. Considering this special case, we implement in our rKKA the possibility to average the CDF obtained for each spectra in a hyperspectral dataset, $\eps_i^p$, after application of the Kramers-Kronig transforms, into an average estimate,

\begin{equation}
 \eps_i^{avg}(E) = \dfrac{1}{N_{p}} \sum_{p=0}^{N_p} \eps_i^p(E) 
\end{equation}

Where $p$ is an index for the spectra in the hyperspectral dataset, running from 0 to $N_p$. Note that in our implementation, $\eps_i^{avg}$ is only used for the calculation of $S_c$; the single $\eps_i^p$ corresponding to each point spectrum are stored and returned as a result after convergence or the last iteration are reached.

\Fig{05} depicts the results of this procedure applied to the processing of an EELS-SL with 21 spectra and a linear thickness gradient from 20 to \SI{120}{\nano\metre}. In this profile, the suppression of spurious contributions is good, and the corrected spectra are in excellent agreement with the expected bulk semi-classical contributions. Moreover, these average rKKA results and the results from single spectrum processing can be compared, see blue dashed and red pointed lines. The average rKKA produces a clearly more reliable reconstruction of the bulk semi-classical term. Most of the Cerenkov-loss signal at the band gap energy onset is correctly modeled and can be subtracted. Meanwhile, the strong rippling at around \SI{15}{\electronvolt} introduced by surface contributions disappears completely. 

This average rKKA reconstruction is useful as long as it incorporates information from different single rKKA thorugh the dataset in an advantageous way. It is possible to quantitatively assess the quality of the presented rKKA reconstructions, since the expected results are known beforehand. We perform this assessment in a logarithmic scale, using the following definition of SNR, 

\begin{equation}
 \snr = 10 \log_{10} \left( \dfrac{\int |S_b^{ELF}| dE}{\int |S-S_c-S_b^{ELF}| dE} \right) 
 \label{eq::snr}
\end{equation}

Where, in the ratio, the denominator contains the integral of the expected semi-classical bulk contribution and the numerator contains the integral of the error for the obtained reconstruction of this contribution. Note that in this process the SNR of a spectrum is indicated by a single value, in dB. \Fig{06} presents such assessment, where the expected and obtained bulk semi-classical contribution are compared for three different panels and through the whole EELS-SL (dashed lines). In the same figure, the quality of the noisy input $S(E)$ is measured by comparing to the noise-less signal (solid lines). Since the quality of the input spectra is dominated by Poisson distributed noise, a linear decay is expected for SNR measured in a logarithmic scale.

\begin{figure}
\includegraphics[width=\linewidth]{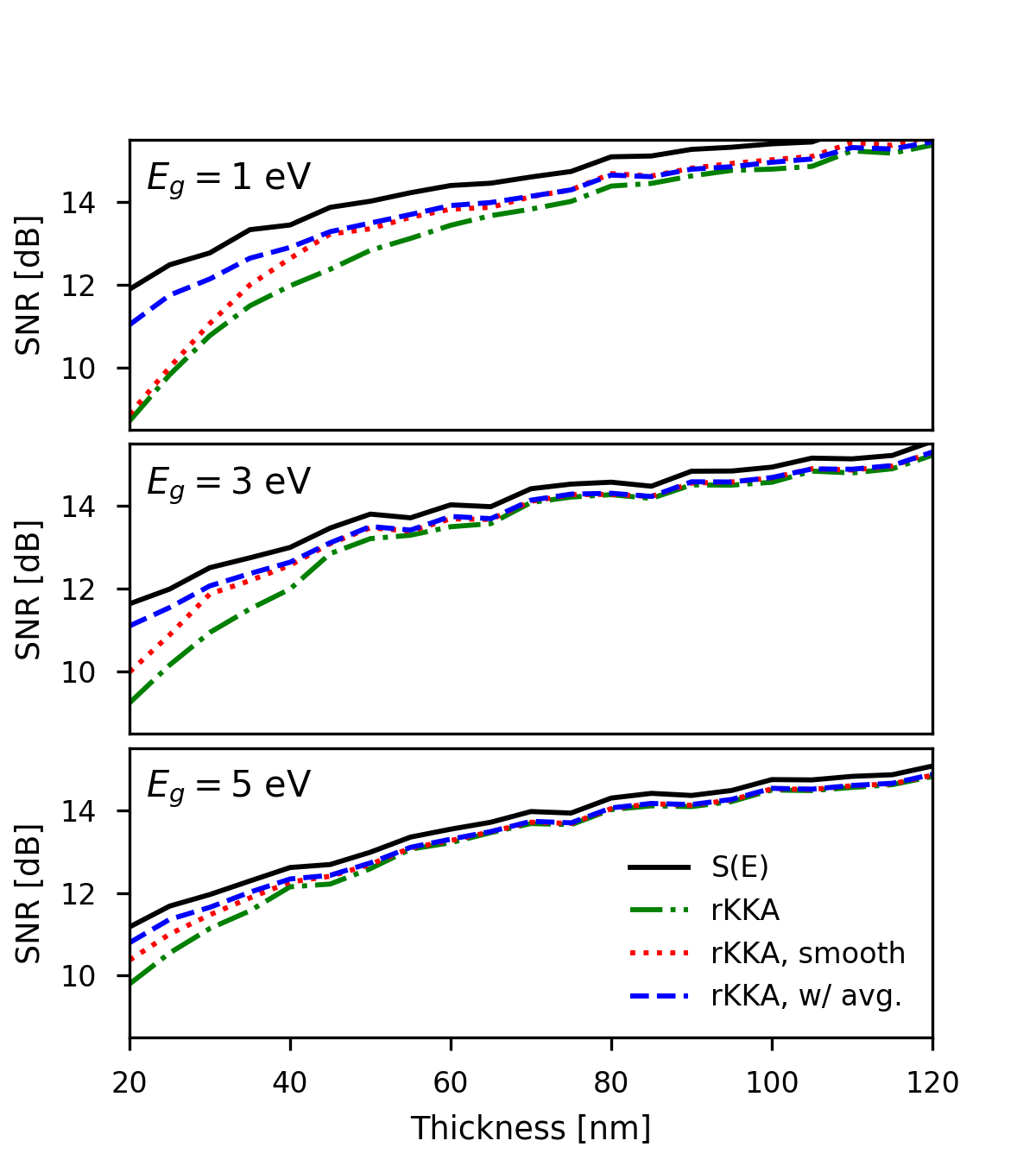}
\caption{ From top to bottom, SNR measured using eq.~\ref{eq::snr} for the rKKA reconstructions of the three $\epstl$ models in \fig{01}, in dashed lines. The rKKA is performed with the naive, regularized and average implementations; as indicated by green, red and blue colors, respectively. The SNR corresponding to the noisy spectra compared to the noise-less case is also included, in solid black lines.
\label{fig::06}}
\end{figure}

Indeed, the thickness dependence of SNR for the noisy input signal is linear, showing a larger SNR than the reconstructions. The single rKKA reconstructions show a slight overall improvement when regularization is used, as rippling features are suppressed or attenuated. The average rKKA shows great improvement in the thinner region, especially for the lower band gap energy cases. In those cases, we have seen that the average results completely remove the spurious features caused by inadequacies in the DF guesses. Also for the average rKKA at the lowest band gap (Eg = 1 eV), the quality drops slightly at the thicker regions. Probably, the origin of this drop are strong features in thin regions not completely eliminated by the averaging.

\begin{figure*}
\includegraphics[width=\linewidth]{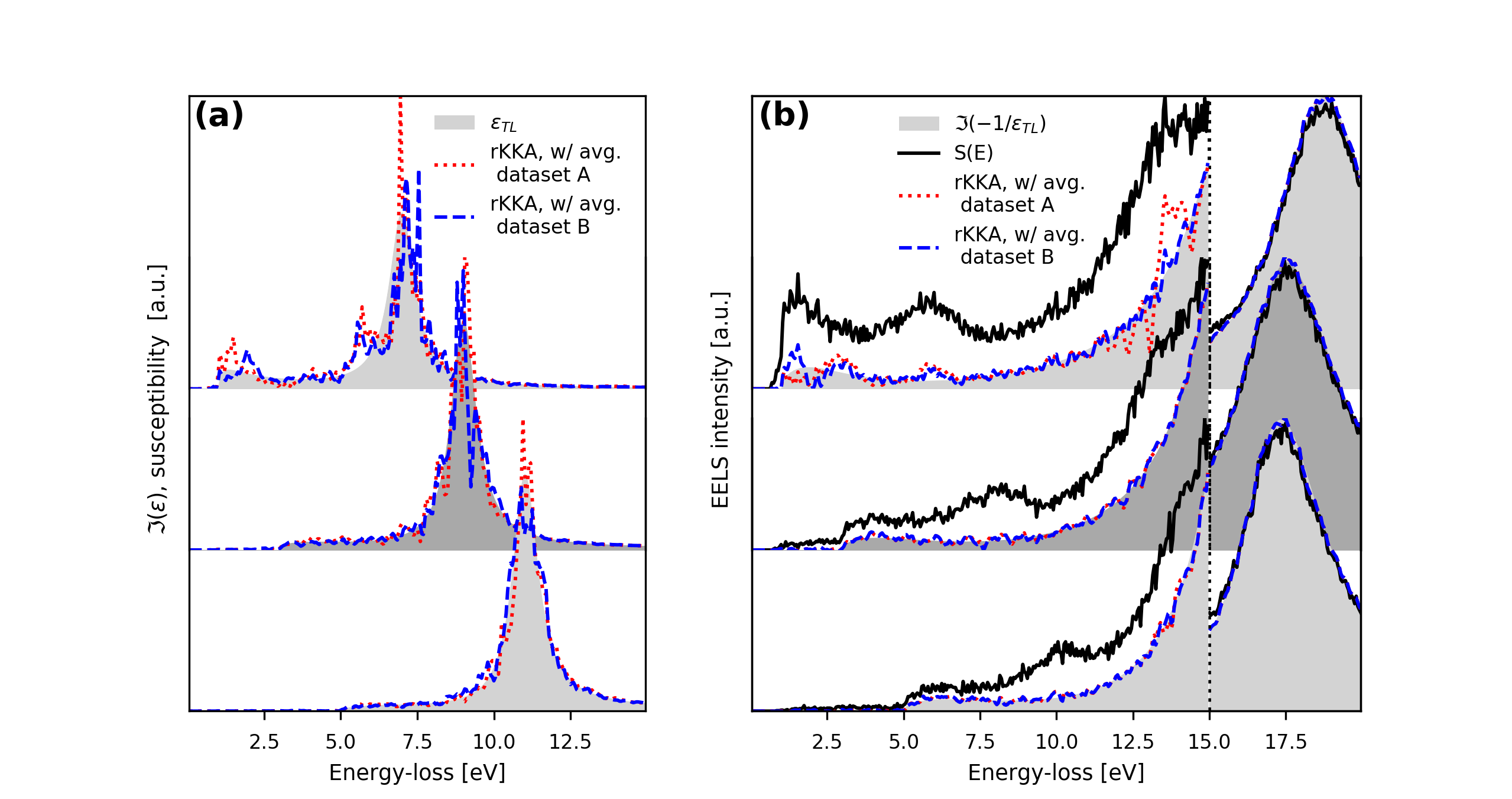}
\caption{Comparison of the $\epstl$ model simulations and average rKKA results for $t=50$ nm and $E_g=1, 3, 5$ eV, from top to bottom (see \fig{01}). Panel (a) shows the imaginary part of the models (grey areas) compared to average rKKA reconstructions obtained using datasets A and B as per explained in the text (red and blue dashed lines, respectively). Panel (b) shows the $\elf$ contribution (grey areas) and the relativistic spectra (solid lines) calculated for these models. Two dashed lines shows the estimated $\elf$ contribution for datasets A and B, again in red and blue color, respectively. \label{fig::07}}
\end{figure*}

We also explore the effect of using a shorter EELS-SL, that contains less spectra in the thicker regions by applying the average rKKA algorithm to two datasets; the first one, labeled A is a short EELS-SL with 21 spectra from 20 to \SI{50}{\nano\metre} (red pointed lines); dataset B, already presented above, has 21 spectra from 20 to \SI{120}{\nano\metre} (blue dashed lines). 

\Fig{07} summarizes results obtained by this procedure at the \SI{50}{\nano\metre} thickness, which is the largest in dataset A. Again for the smaller band gap energy, the rippling features of the thinner regions have been introduced into the result for this thickness. The amount of spectra in this shorter EELS-SL at the thicker region is not sufficient to compensate for the errors introduced in the thinner regions. 

Nevertheless, comparison of Figs.~\ref{fig::01},\ref{fig::04} and \ref{fig::05} shows that average rKKA results in all other cases are better than single rKKA results. Averaging adds to the robustness of the algorithm, given that a sufficient number of spectra are acquired from regions not critically impacted by spurious contributions (thinner or thicker). 

\section{Conclusion \label{sec::end}}
In the case that a relativistic contribution to EELS cannot be disregarded traditional KKA does not guarantee retrieving the correct DF, even from perfect noise-less input data. In turn, rKKA allows to retrieve the correct CDF and a meaningful correction term, even in the naive implementation. However, in order to use low-loss EELS for standard-free measurement of the dielectric properties, the speed and reliability of the rKKA present challenging issues. In this paper, we have explored and proposed solutions to these issues.

The time-consuming and error prone computation of relativistic spectra is one of the main issues. According to our calculations, the optimized numerical integration scheme using Simpson-rule improves one order of magnitude the speed of the DDCS integration. This feat additionally allows batch processing of hyperspectral datasets, which becomes relevant when analyzing noisy experimental data.

The results from the naive implementation of rKKA are plagued with artifacts, related to the inaccuracy of the initial guess for the CDF and the noise-response of the DDCS integration. When treated using simple regularization by bounding and smoothing, we have showed, these errors could be many times suppressed or at least attenuated. This methodology makes rKKA more robust in the majority of cases. However, in very thin samples regularization is by itself insufficient, and some errors always remain. We have proposed to use batch analysis of hyperspectral datasets, showing how averaging of the CDF improves the performance of rKKA. We foresee that this simple averaging trick could be improved in the future by weighting in differently the information from different energy-loss regions according to the thickness.

The present study broadens the application range of KKA to situations in which relativistic and surface losses have larger impact in the spectra than ever before. Some limitations of the technique remain, the application of KKA to very thin or thick specimens still remains problematic because of the inadequacy of the normalization procedure and the effect of beam broadening.

%\appendix
%\beginsupplement

\bibliography{refs}

\end{document}